\documentclass[review,12pt]{elsarticle}
\usepackage{amssymb}
\usepackage{amsthm}
\usepackage{amsmath}
\usepackage{mathrsfs}
\usepackage{graphicx}
\usepackage{epstopdf}
\usepackage{float}
\usepackage{caption}
\usepackage{subcaption}
\usepackage{bm}
\usepackage{bbm}
\usepackage{mathrsfs}
\usepackage{cleveref}
\usepackage{soul}
\usepackage{multirow}
\usepackage{xcolor}
\usepackage{siunitx}
\usepackage[margin=2cm]{geometry}% TO PLAY WITH THE MARGINS
\usepackage{framed} % To put a box to the nomenclature
\usepackage{nomencl} %to add the nomenclature
\makenomenclature
\biboptions{sort&compress}

\journal{ }
\makeatletter
\def\@author#1{\g@addto@macro\elsauthors{\normalsize%
    \def\baselinestretch{1}%
    \upshape\authorsep#1\unskip\textsuperscript{%
      \ifx\@fnmark\@empty\else\unskip\sep\@fnmark\let\sep=,\fi
      \ifx\@corref\@empty\else\unskip\sep\@corref\let\sep=,\fi
      }%
    \def\authorsep{\unskip,\space}%
    \global\let\@fnmark\@empty
    \global\let\@corref\@empty  %% Added
    \global\let\sep\@empty}%
    \@eadauthor={#1}
}
\makeatother

\begin{document}

\begin{frontmatter}



\title{Cohesive zone modelling of hydrogen assisted fatigue crack growth: the role of trapping}


\author{Rebeca Fern\'{a}ndez-Sousa \fnref{Uniovi}}

\author{Covadonga Beteg\'{o}n \fnref{Uniovi}}

\author{Emilio Mart\'{\i}nez-Pa\~neda\corref{cor1}\fnref{IC}}
\ead{e.martinez-paneda@imperial.ac.uk}

\address[Uniovi]{Department of Construction and Manufacturing Engineering, University of Oviedo, Gij\'{o}n 33203, Spain}

\address[IC]{Department of Civil and Environmental Engineering, Imperial College London, London SW7 2AZ, UK}

\cortext[cor1]{Corresponding author.}

\begin{abstract}
We investigate the influence of microstructural traps in hydrogen-assisted fatigue crack growth. To this end, a new formulation combining multi-trap stress-assisted diffusion, mechanism-based strain gradient plasticity and a hydrogen- and fatigue-dependent cohesive zone model is presented and numerically implemented. The results show that the ratio of loading frequency to effective diffusivity governs fatigue crack growth behaviour. Increasing the density of \emph{beneficial} traps, not involved in the fracture process, results in lower fatigue crack growth rates. The combinations of loading frequency and carbide trap densities that minimise embrittlement susceptibility are identified, providing the foundation for a rational design of hydrogen-resistant alloys.\\ 

%% INSTRUCTIONS IJHE 
%A concise and factual abstract is required. The abstract should be written in the present tense and be no longer than 150 words maximum. It should briefly state the purpose of the research, the principal results and major conclusions.
%% INSTRUCTIONS INT J FATIGUE
%A concise and factual abstract of approximately 100 words is required. The abstract should state briefly the purpose of the research, the principal results and major conclusions. 
\end{abstract}


\end{frontmatter}


% To create the nomenclature in TexMaker go to Tools, open terminal and then type:
% makeindex <filename>.nlo -s nomencl.ist -o <filename>.nls
% i.e., in this case: makeindex Draft1nomarking.nlo -s nomencl.ist -o Draft1nomarking.nls
% This will create an .nls file that has to be uploaded to arXiv

\begin{framed}
\nomenclature{$a$}{crack length}
\nomenclature{$b$}{Burgers vector}
\nomenclature{$C^*, \, m$}{Paris law coefficients}
\nomenclature{$C$}{total hydrogen concentration}
\nomenclature{$C_L, \, C_T$}{hydrogen concentration in lattice and trapping sites}
\nomenclature{$C_{L,0}$}{initial lattice hydrogen concentration}
\nomenclature{$D_L, D_e$}{lattice and effective diffusion coefficients}
\nomenclature{$D, \, D_c, \, D_m$}{damage variable: total, cyclic and monotonic}
\nomenclature{$\overline{da/dN}$}{normalised fatigue crack growth rate}
\nomenclature{$E$}{Young's modulus}
\nomenclature{$f$}{load frequency}
\nomenclature{$K_T^{(i)}$}{equilibrium constant for the $i$th type of trapping sites}
\nomenclature{$\Delta K$}{stress intensity factor range}
\nomenclature{$K_{min}, \, K_m, \, K_{max}$}{minimum, mean and maximum stress intensity factor}
\nomenclature{$K_0$}{reference stress intensity factor}
\nomenclature{$\ell$}{material gradient length scale}
\nomenclature{$M$}{Taylor's factor}
\nomenclature{$N$}{number of cycles}
\nomenclature{$n$}{strain hardening exponent}
\nomenclature{$N_A$}{Avogadro's number}
\nomenclature{$N_L$}{number of lattice sites per unit volume}
\nomenclature{$N_T^{(i)}$}{number of sites per unit volume for the $i$th type of trapping sites}
\nomenclature{$\mathcal{R}$}{universal gas constant}
\nomenclature{$R$}{load ratio}
\nomenclature{$r_0$}{initial crack tip blunting radius}
\nomenclature{$\bar{r}$}{Nye's factor}
\nomenclature{$r$,\, $\theta$}{polar coordinates}
\nomenclature{$T$}{absolute temperature}
\nomenclature{$T_n$}{normal cohesive traction}
\nomenclature{$t$}{time}
\nomenclature{$u_x , \, u_y$}{horizontal and vertical components of the displacement field}
\nomenclature{$\bar{V}_H$}{partial molar volume of hydrogen}
\nomenclature{$V_M$}{molar volume of the host lattice}
\nomenclature{$W_B^{(i)}$}{binding energy for the $i$th type of trapping sites}
\nomenclature{$\beta$}{number of lattice sites per solvent atom}
\nomenclature{$\Delta_n$}{normal cohesive separation}
\nomenclature{$\delta_n$}{characteristic normal cohesive length}
\nomenclature{$\delta_{\Sigma}$}{accumulated cohesive length}
\nomenclature{$\varepsilon^p$}{equivalent plastic strain}
\nomenclature{$\eta^p$}{effective plastic strain gradient}
\nomenclature{$\theta_L, \, \theta_T^{(i)}$}{occupancy of lattice and $i$th type of trapping sites}
\nomenclature{$\mu$}{shear modulus}
\nomenclature{$\nu$}{Poisson's ratio}
\nomenclature{$\rho$}{dislocation density}
\nomenclature{$\rho_S$}{statistically stored dislocation (SSD) density}
\nomenclature{$\rho_G$}{geometrically necessary dislocation (GND) density}
\nomenclature{$\sigma_\Sigma$}{cohesive endurance limit}
\nomenclature{$\sigma_{max}, \, \sigma_{max,C}, \, \sigma_{max,0}$}{current, hydrogen-degraded and initial cohesive strengths}
\nomenclature{$\sigma_H$}{hydrostatic stress}
\nomenclature{$\sigma_f$}{tensile flow stress}
\nomenclature{$\sigma_y$}{initial yield stress}
\nomenclature{$\tau$}{shear flow stress}
\nomenclature{$\phi_n$}{normal cohesive energy}
\printnomenclature
\end{framed}

\section{Introduction}
\label{Sec:Intro}
% INSTRUCTIONS IJHE
% From the guide to authors: The length should not normally exceed 8000 words and 12 diagrams - this corresponds to approximately 12 journal pages. That is roughly 30 pages in this format. 
% INSTRUCTIONS INT. J FATIGUE
%Original high-quality research papers (no more than 30 double spaced pages, written using Times New Roman in 10-12 point size font, including figures and references)

When exposed to hydrogen, metallic materials experience a significant loss of ductility, toughness and fatigue crack growth resistance \cite{Gangloff2003,Djukic2019}. This phenomenon, termed hydrogen embrittlement, is arguably the biggest threat to the deployment of a hydrogen energy infrastructure and the cause of numerous structural integrity problems in the transport, defence, marine and construction sectors \cite{RILEM2021}. Most often, susceptible components are subjected to cyclic loads, and this has triggered significant interest in understanding the interplay between hydrogen and fatigue damage (see, e.g., \cite{Martin2013,Colombo2015,Yamabe2017,Castelluccio2018,Ogawa2020} and Refs. therein).\\

Hydrogen ingress into a metal can occur during manufacturing operations, such as casting, welding, machining or electroplating, and through exposure to hydrogenous environments such as water vapour, aqueous electrolytes or hydrogen-containing gas. Following ingress, atomic hydrogen diffuses through the crystal lattice and resides at either interstitial lattice sites or microstructural trapping sites (e.g., dislocations, grain boundaries, voids, carbides and interfaces). Whether embrittlement is governed by the hydrogen content in lattice or by the one in trap sites is still a matter of debate \cite{Ayas2014,Harris2018,Shishvan2020,Anand2019,IJP2021}, and trapping characteristics vary from one material to another \cite{Ai2013,Barrera2018,IJF2020}. However, the hydrogen concentration in lattice sites $C_L$ is generally in equilibrium with the hydrogen concentration in trapping sites $C_T$ \cite{Oriani1974}; implying that there is a unique relationship between them, and that the accumulation of trapped hydrogen follows that of lattice hydrogen. Hence, an accurate characterization of lattice hydrogen diffusion is of utmost importance. Experiments show that the degree of embrittlement is sensitive to the hydrogen content and the loading rate, with the limiting cases being given by sufficiently fast tests (where hydrogen transport is negligible) and by sufficiently slow tests (where hydrogen transport has reached the steady state) \cite{Momotani2017}. In fatigue experiments, this results in a sensitivity to the loading frequency. Fatigue crack growth rates increase with decreasing frequency as there is more time for the hydrogen to accumulate in the fracture process zone \cite{Gangloff1990,Gangloff2012}. And again, the behaviour is bounded between two limiting cases: sufficiently high and low loading frequencies. In fact, some experiments show that if the loading frequency is sufficiently high, then embrittlement is precluded and fatigue crack growth rates become comparable to those measured in inert environments \cite{Murakami2010a,Fassina2013,Tazoe2017,Alvaro2019,Peral2019}.\\

Recently, Fern\'{a}ndez-Sousa \textit{et al.} \cite{AM2020} have demonstrated that the fatigue behaviour is governed by the ratio between the loading frequency ($f$) and the material effective diffusion coefficient ($D_e$). Their numerical results showed that the maximum hydrogen content attained in the fracture region was sensitive to $f/D_e$ for both open systems (where there is a permanent source of hydrogen) and closed-systems (where the content of hydrogen is limited). Their findings imply that the hydrogen content can be decreased below the embrittlement threshold if $D_e$ is reduced, suggesting that materials can be engineered to bring down their hydrogen diffusivity and susceptibility to hydrogen-assisted fatigue crack growth. The effective diffusivity of materials can be reduced by increasing the density of \emph{beneficial traps} - microstructural trapping sites that are not involved in the fracture process \cite{Ramjaun2018,Turk2018}. Fern\'{a}ndez-Sousa \textit{et al.} \cite{AM2020} showed that increasing the density of carbides by 3 orders of magnitude in a CrMo steel enabled extending by an order of magnitude the regime of safe frequencies at which hydrogen has no effect. However, the analysis of Fern\'{a}ndez-Sousa \textit{et al.} \cite{AM2020} was based on the assumption of a critical hydrogen threshold, without explicitly simulating crack growth.\\

In this work, we combine a coupled deformation-diffusion multi-trap model with a cohesive zone formulation for cyclic damage to simulate hydrogen-assisted fatigue crack growth and investigate the role of microstructural traps. Very few works have been published reporting fatigue crack growth predictions in the presence of hydrogen. Moriconi \textit{et al.} \cite{Moriconi2014} used an irreversible cohesive zone model to investigate the fatigue resistance of a 15-5PH martensitic steel intended for gaseous hydrogen storage. Del Busto \textit{et al.} \cite{EFM2017} also combined a fatigue cohesive zone model with a stress-assisted diffusion formulation, quantifying the influence of the loading frequency and mapping the resulting regimes. Very recently, Golahmar \textit{et al.} \cite{Golahmar2022} presented the first phase field formulation for hydrogen-assisted fatigue. All of these works were limited to one trap type, at most, and did not explore the influence of increasing the trap density. Also, conventional continuum models (such as $J_2$ plasticity or linear elasticity) were used to predict material deformation. However, it has been shown that conventional plasticity theory fails to capture the crack tip stress elevation associated with non-uniform plastic deformation and Geometrically Necessary Dislocations (GNDs) \cite{Wei1997,Komaragiri2008,IJP2016}. This is particularly relevant in hydrogen embrittlement as hydrogen accumulates in areas of high hydrostatic stress; gradient-enhanced calculations reveal very large hydrogen concentrations within the critical distance of hydrogen-assisted cracking, rationalising mechanisms such as hydrogen-enhanced decohesion \cite{IJHE2016,AM2016,JMPS2020}. Also, GNDs act as trapping sites and their density becomes significant close to cracks or other stress concentrators. In this work, we use a mechanism-based strain gradient plasticity model to capture the role of GNDs and the associated crack tip stress elevation \cite{Gao1999,CM2017}. Thus, the present work also provides the first theoretical and computational framework to model hydrogen-assisted fatigue crack growth accounting for the role of GNDs and plastic strain gradients.\\

The remainder of this manuscript is organised as follows. The theoretical framework combining multi-trap stress-assisted hydrogen diffusion, mechanism-based strain gradient plasticity and a hydrogen- and cyclic-damage cohesive zone model is presented in Section \ref{Sec:Theory}. Details of the material investigated, the numerical implementation, and the boundary value problem used are given in Section \ref{Sec:Met}. The results are then presented in Section \ref{Sec:Results}. First, the influence on fatigue crack growth rates of varying the trap density is quantified. Secondly, the sensitivity to the initial hydrogen content is investigated. Thirdly, we conduct numerical experiments at different loading frequencies to study the interplay between frequency and diffusivity. Finally, maps are built to assist in the development of hydrogen-resistant alloys that exploit the concept of beneficial traps to reduce or suppress the susceptibility to hydrogen embrittlement over technologically-relevant loading frequencies. Concluding remarks end the manuscript in Section \ref{Sec:ConcludingRemarks}.

\section{Theory}
\label{Sec:Theory}

\subsection{A multi-trap model for hydrogen diffusion}
\label{Sec:hydrogenTransportModel}

The total hydrogen concentration is given by the sum of the hydrogen concentration at interstitial lattice sites and the hydrogen concentration at microstructural trapping sites, $C=C_L+C_T$. The hydrogen concentration of lattice sites is related to the lattice site occupancy fraction $\theta_L$ and the number of lattice sites per unit volume $N_L$, as follows
\begin{equation}\label{eq:CL}
    C_L=\theta_{L} N_L \, , \,\,\,\,\,\, \text{with} \,\,\,\,\,\, N_L = \frac{\beta N_A \rho_M}{M_M} \, .
\end{equation}

\noindent Here, $\beta$ is the number of interestitial sites per solvent atom, $N_A$ is Avogadro's number, $M_M$ is the atomic weight and $\rho_M$ the density. In iron-based bcc materials $\beta=6$ \cite{Krom1999}, $\rho_M=7870$ kg/m$^3$ and $M_M=55.8 \times 10^{-3}$ kg/mol, giving $N_L=5.1 \times 10^{29}$ sites/m$^3$. Similarly, the hydrogen concentration for the $i$th type of trapping sites is given by
\begin{equation}\label{eq:CT}
    C_T^{(i)} = \theta_T^{(i)} N_T^{(i)},
\end{equation}

\noindent where $\theta_T$ is the trap occupancy and $N_T$ is the trap density. For most trap types, $N_T$ is a material property that remains constant throughout the duration of the experiment. However, the number of dislocation trap sites per unit volume evolves with the dislocation density; a Taylor-based formulation is used in this work to determine the dislocation density from the mesoscale notions of plastic strains and plastic strain gradients.\\

Thermodynamic equilibrium between traps and interstitial sites is assumed, following the work by Oriani \cite{Oriani1974}. Accordingly, the following Fermi-Dirac relation between the occupancy of the $i$th type of trapping sites and the fraction of occupied lattice sites is adopted
\begin{equation}\label{eq:Oriani}
    \frac{\theta_T^{(i)}}{1 - \theta_T^{(i)}} = \frac{\theta_L}{1- \theta_L} \exp \left( \frac{-W_B^{(i)}}{\mathcal{R}T} \right),
\end{equation}

\noindent with $W_B^{(i)}$ being the trap binding energy for the $i$th type of trap, $\mathcal{R}=8.3145$ J/(mol$\cdot$K) the universal gas constant and $T$ the absolute temperature. This relation enables formulating the hydrogen transport equation solely in terms of the lattice hydrogen concentration. For a partial molar volume of hydrogen in solid solution $\bar{V}_H$, a hydrostatic stress $\sigma_H$, and a lattice diffusion coefficient $D_L$, the mass transport equation reads
\begin{equation}
        \frac{D_L}{D_e}\frac{dC_L}{dt}=D_L \nabla^2 C_L-\nabla \left( \frac{D_L C_L}{\mathcal{R}T} \bar{V}_H \nabla \sigma_H  \right ) ,
    \label{Dif}
\end{equation}

\noindent with the effective diffusion coefficient $D_e$ being defined as
\begin{equation}\label{eq:De}
    D_e = D_L \frac{C_L}{C_L+ \sum_i C_T^{(i)} \left( 1 - \theta_T^{(i)} \right)} \, .
\end{equation}

\subsection{Mechanism-based strain gradient plasticity}
\label{Sec:CMSG}

The mechanical behaviour of the solid is characterised by means of the so-called mechanism-based strain gradient (MSG) plasticity theory \cite{Gao1999,Huang2004a}. MSG plasticity is based on Taylor's dislocation model \cite{Taylor1938} and thus provides an enriched continuum description capable of capturing the role of GNDs in elevating crack tip stresses and hydrogen concentration, as well as the evolution of the total dislocation density, as required for quantifying the dislocation trap density $N_T^{(d)}$.\\

In Taylor's dislocation model \cite{Taylor1938}, the shear flow stress $\tau$ is estimated from the shear modulus $\mu$, the Burgers vector $b$ and the total dislocation density $\rho$ as,
\begin{equation}
    \tau = 0.5 \mu b \sqrt{\rho} \, .
\end{equation}

The tensile flow stress $\sigma_f$ is then related to $\tau$ through the Taylor factor $M$. Considering that the total dislocation density can be additively decomposed into the GND density, $\rho_{G}$, and the density of statistically stored dislocations (SSDs), $\rho_{S}$, the flow stress can be expressed as
\begin{equation}\label{eq:sigma_f}
    \sigma_{f} = M \tau = 0.5 M \mu b \sqrt{\rho_S + \rho_G} \, ,
\end{equation}

\noindent where $b$ and $M$ respectively equal 0.2725 nm and 2.9 for bcc metals. The GND density is defined in terms of Nye's factor $\bar{r}$ ($\approx 1.9$ \cite{Arsenlis1999}), Burger's vector, and the effective plastic strain gradient $\eta^p$ as
\begin{equation}\label{eq:rhoG}
    \rho_{G} = \bar{r} \frac{\eta^p}{b} \, .
\end{equation}

\noindent Here, the effective plastic strain gradient is computed from the plastic strain tensor as,
\begin{equation}
    \eta^p = \sqrt{\frac{1}{4} \eta_{ijk}^p \eta_{ijk}^p} \,\,\,\,\,\,\, \text{with} \,\,\,\,\,\,\, \eta_{ijk}^p=\varepsilon_{ik,j}^p+\varepsilon_{jk,i}^p-\varepsilon_{ij,k}^p
\end{equation}

Combining (\ref{eq:sigma_f}) and (\ref{eq:rhoG}), the SSD density $\rho_{S}$ can be determined knowing the relation in uniaxial tension $(\eta = 0)$ between the flow stress and the material stress-strain curve as follows
\begin{equation}\label{eq:rhoS}
    \rho_{S} = \left( \frac{\sigma_{ref} f \left( \varepsilon^p \right)}{0.5 M \mu b} \right)^2 \, ,
\end{equation}

\noindent where $\sigma_{ref}$ is a reference stress and $f(\varepsilon^p)$ is a non-dimensional function of the equivalent plastic strain, as given by the uniaxial stress-strain curve. Substituting back into (\ref{eq:sigma_f}), the flow stress can be re-formulated as
\begin{equation}\label{eq:sF_msg}
 \sigma_f = \sigma_{ref} \sqrt{f^2 \left( \varepsilon^p \right) + \ell \eta^p} \, .
\end{equation}

Eq. (\ref{eq:sF_msg}) introduces a plastic length scale $\ell$, which naturally accounts for the material's length-scale dependency arising from the Burgers vector. Conventional von Mises plasticity is recovered when the characteristic length of plastic deformation outweighs the GNDs-related term $\ell \eta^p$, where $\ell$ is a material parameter characterising the capacity to undergo hardening due to the presence of GNDs. The magnitude of $\ell$ can be determined by fitting the size-dependent response measured in micro-scale experiments, with typical values for $\ell$ ranging between 1 and 10 \si{\micro\metre} \cite{IJES2020}. Accordingly, a value of $\ell=5$ \si{\micro\metre} is considered in this work. 

\subsection{Cohesive zone model}
A cohesive zone formulation is employed to model crack growth accounting for the combined effects of hydrogen and fatigue damage. The model is an extension of the cohesive zone formulation for fatigue developed by Roe and Siegmund \cite{Roe2003}, incorporating a new phenomenological hydrogen degradation law. The focus is on mode I conditions and therefore only the normal components of the critical variables are presented.\\

The constitutive behavior of the cohesive interface is based on the exponential traction-separation law first presented by Xu and Needleman \cite{Xu1993}. Thus, the relation between the normal traction ($T_n$) and the associated displacement jump ($\Delta_n$) is given by
\begin{equation}\label{eq:Tn}
    T_n=\frac{\phi_n}{\delta_n}\left( \frac{\Delta_n}{\delta_n} \right)\exp\left( -\frac{\Delta_n}{\delta_n} \right) \, ,
\end{equation}

\noindent where $\delta_n$ is the characteristic cohesive length under normal separation and $\phi_n$ is the fracture energy. The latter is defined as follows
\begin{equation}\label{eq:Sep}
    \phi_n=\exp(1) \, \sigma_{max,0} \, \delta_n ,
\end{equation}

\noindent where $\sigma_{max,0}$ is the initial cohesive strength, before the hydrogen and fatigue degradations are accounted for.\\

\begin{figure}[H]
  \makebox[\textwidth][c]{\includegraphics[width=0.7\textwidth]{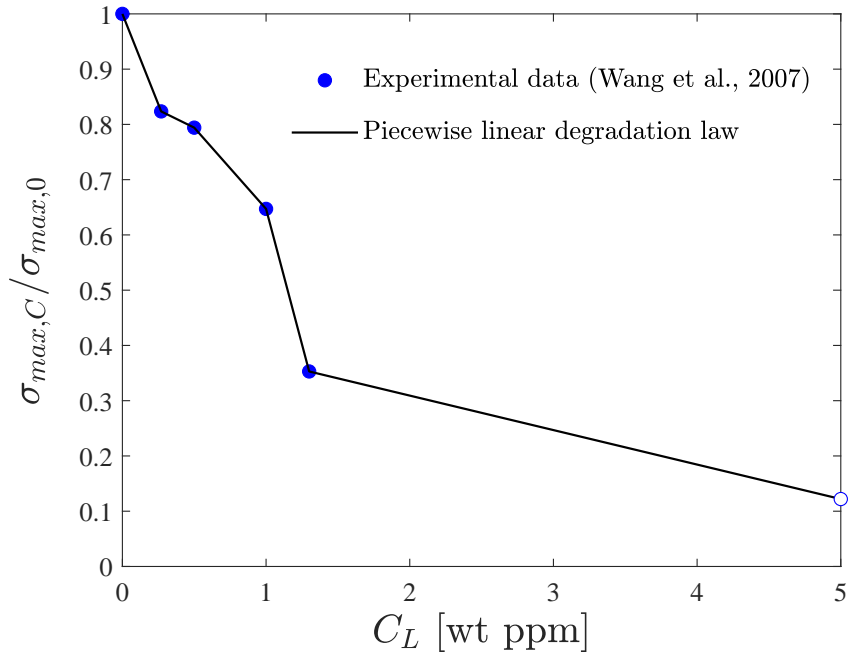}}
  \caption{Sensitivity of the material strength to the hydrogen content. Phenomenological degradation law based on the experiments by Wang \textit{et al.} \cite{Wang2007} on AISI 4135 steel. The material strength is given normalised by the material strength in air. As explained in the text, a saturation point is provided to extend the law to high hydrogen concentrations.}
  \label{fig:LeyH}
\end{figure}

The role of hydrogen in degrading the fracture energy of the solid is incorporated in a phenomenological way, based on the experiments by Wang \textit{et al.} \cite{Wang2007} (see also Ref. \cite{Yu2016a}). Fracture energy degradation laws based on first principles have been proposed \cite{Serebrinsky2004,CMAME2018} and have significant appeal, as they would ultimately enable predictions without the need for experimental calibration. However, this approach requires defining \textit{a priori} the failure mechanism, as well as the nature of the decohering interface and its binding energy. To retain generality, we choose instead to base our analysis on an experimentally inferred degradation law. The influence of hydrogen is incorporated \textit{via} the cohesive strength, as atomistic calculations show that, unlike $\sigma_{max}$, the critical separation is rather insensitive to the hydrogen coverage \cite{VanderVen2003}. Wang \textit{et al.} \cite{Wang2007} conducted uniaxial tension tests on smooth bars of AISI 4135 steel for different hydrogen concentrations. Their results are shown in Fig. \ref{fig:LeyH} using symbols and normalising the measured strength by the tensile strength in air. This enables establishing a quantitative, piece-wise linear relationship between the initial cohesive strength and the hydrogen-degraded cohesive strength,
\begin{equation}
    \sigma_{max,C} = f (C_L) \sigma_{max,0}
\end{equation}

\begin{figure}[H]
  \makebox[\textwidth][c]{\includegraphics[width=0.9\textwidth]{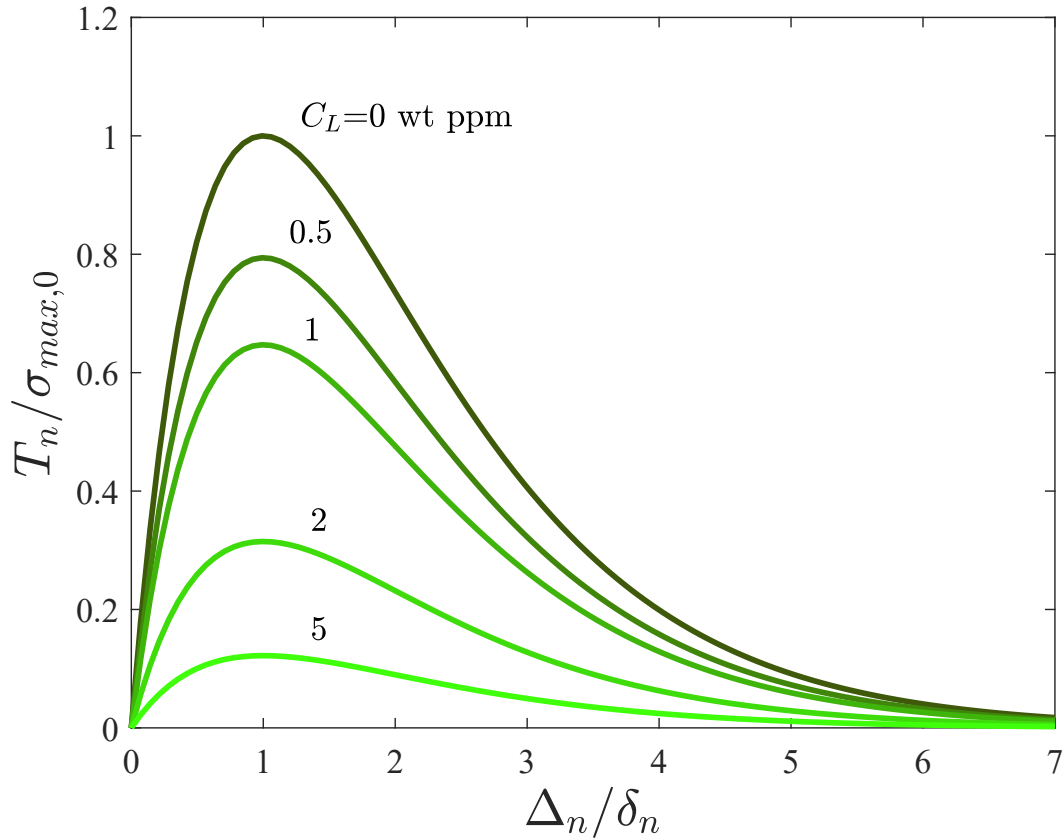}}
  \caption{Traction-separation law characterising the cohesive zone model for various hydrogen concentrations (in wt ppm).}
  \label{fig:Cohesivos}
\end{figure}

It should be noted that Wang \textit{et al.} \cite{Wang2007} only carried out experiments until reaching a hydrogen concentration of 1.3 wt ppm. To provide predictions for larger hydrogen contents (as attained locally in the vicinity of cracks), we add an additional data point at 5 wt ppm and $\sigma_{max,C}=0.122\sigma_{max,0}$. These choices aim at reducing the slope to capture the commonly-observed saturation effect and avoid overestimating the degree of embrittlement; the saturated cohesive strength is taken as the one resulting from the quantum-mechanical law by Serebrinsky \textit{et al.} \cite{Serebrinsky2004} under conditions of full hydrogen coverage. For hydrogen concentrations beyond the saturation point, no further degradation of the cohesive strength is assumed. The sensitivity of the traction-separation law to the hydrogen concentration is illustrated in Fig. \ref{fig:Cohesivos}. 

Finally, the influence of fatigue damage is incorporated. Following the model of Roe and Siegmund \cite{Roe2003}, a damage mechanics approach is adopted by which the effective cohesive strength is degraded using a damage variable $D$; that is
\begin{equation}\label{eq:max}
    \sigma_{max}=\sigma_{max,C}\left(1-D\right) \, .
\end{equation}

Damage due to both monotonic and cyclic loading must be captured, and as a result the damage state is defined as the maximum of these two contributions:
\begin{equation}
    D=\int_0^t \max \left( \dot{D}_c,\dot{D}_m \right) \, \text{d}t \, ,
\end{equation}

\noindent where $\dot{D}_c$ and $\dot{D}_m$ respectively denote the cyclic and monotonic damage rates. The latter is only updated when the largest stored value of $\Delta_n$ is greater than $\delta_n$ and is given by,
\begin{equation}\label{eq:Daño_mon}
   \dot{D}_m=\frac{\max \left( \Delta_n \right)|_{t_i} - \max \left( \Delta_n  \right)|_{t_{i-1}}}{4\delta_n} \, ,
\end{equation}

\noindent where $t_i$ denotes the current time increment and $t_{i-1}$ the previous one. Finally, fatigue is characterised by the following damage evolution law,
\begin{equation}\label{eq:Daño_ciclico}
    \dot{D}_c=\frac{|\dot{\Delta}_n|}{\delta_{\Sigma}}\left[\frac{T_n}{\sigma_{max,C}}-\frac{\sigma_\Sigma}{\sigma_{max,0}}\right]H\left(\bar{\Delta}_n - \delta_n\right) \, , \,\,\,\,\,\,\, \text{with} \,\,\,\, \bar{\Delta}_n=\int_0^t |\dot{\Delta}_n| \, \text{d}t \, .
\end{equation}

\noindent Here, $\sigma_\Sigma$ is the cohesive endurance limit, $H$ is the Heaviside function and $\delta_{\Sigma}$ is the accumulated cohesive length. Thus, Eq. (\ref{eq:Daño_ciclico}) incorporates key characteristics of continuum damage mechanics laws \cite{Siegmund2004}: (i) damage starts if a deformation measure is greater than a critical magnitude, as determined by the Heaviside function; (ii) the damage rate is related to the increment of deformation and the current load level; and (iii) a stress endurance limit exists, as given by $\sigma_f$, below which cyclic loading can proceed infinitely without failure. Here, following Ref. \cite{Roe2003}, we assume that $\delta_\Sigma=4 \delta_n$ and $\sigma_\Sigma/\sigma_{max,C}=0.25$. 

\section{Methodology}
\label{Sec:Met}

\subsection{Material properties}
\label{Sec:Material}

Our numerical experiments are conducted on a AISI 4140 steel that has been extensively characterised \cite{AM2020,Peral2019,Zafra2018}, both in terms of its trapping characteristics and of its fatigue behaviour in the presence of hydrogen. Fatigue experiments were conducted at different loading frequencies showing that hydrogen had no effect on fatigue crack growth rates if a sufficiently high loading frequency was used \cite{AM2020,Zafra2018}. This AISI 4140 (42CrMo4) steel was austenitized at 845$^\circ$C for 40 min, quenched in water, and tempered at 700$^\circ$C for two hours. The elastic properties are given by a Young's modulus of $E=220$ GPa and a Poisson's ratio of $\nu=0.3$. The plastic behaviour is captured by the following hardening power law:
\begin{equation}
    \sigma = \sigma_y \left( 1 + \frac{E \varepsilon^p}{\sigma_y} \right)^{(1/n)} \, ,
\end{equation}

\noindent with the yield stress being $\sigma_y=622$ MPa and the strain hardening coefficient $n=10$. Note that the reference stress in Eq. (\ref{eq:sF_msg}) corresponds to $\sigma_{ref}=\sigma_y (E / \sigma_y)^{(1/n)}$ while $f \left( \varepsilon^p \right)= \left( \varepsilon^p + \sigma_y / E \right)^{(1/n)}$. Thus, hardening is assumed to be purely isotropic, neglecting kinematic hardening effects. However, kinematic hardening can play a role; even during static (monotonic) fracture, kinematic hardening impacts crack growth resistance due to non-proportional straining \cite{JAM2018}. Nonetheless, these effects are likely to be relatively insensitive to the trap density and, given the conditions considered here (load ratio of $R=0.1$, a small number of cycles, and short crack extensions), arguably of secondary importance. The fracture behaviour is characterised by an initial cohesive strength of $\sigma_{max,0}=4\sigma_y$ \cite{Tvergaard1992,JMPS2019}.\\

Regarding the hydrogen transport properties, the lattice diffusion coefficient has been found to be $D_L=1.3 \times 10^{-9}$ m$^2$/s \cite{Peral2019} and the initial lattice hydrogen concentration after pre-charging was estimated to be equal to $C_{L,0}=1.06$ wt ppm by combining Thermal Desorption Spectroscopy (TDS) and diffusion modelling \cite{AM2020}. The partial molar volume of hydrogen in iron-based materials is taken to be $\bar{V}_H=2 \times 10^{-6}$ m$^3$/mol \cite{Hirth1980}. As elaborated in Ref. \cite{AM2020}, three trap types have been identified: dislocations, carbides and martensitic interfaces. The estimated binding energies and trap densities are given in Table \ref{tab:energy}, with the trap density for dislocations being that of the unstressed state ($N_{T,0}^{(d)}$).   

\begin{table}[H]
  \centering
  \caption{Binding energies $W_B$ and trap densities $N_T$ for each trap type in the AISI 4140 steel considered \cite{AM2020,Zafra2020}.}
    \begin{tabular}{|c|c|c|}
    \hline
    \textbf{Trap type} & $W_B$ [kJ/mol] & \multicolumn{1}{c|}{$N_T$ [sites/m$^{3}$]} \\
    \hline
    Dislocations & -35.2  & $4.93 \times 10^{23}$ \\
    Carbides & -21.4  & $3.61 \times 10^{23}$ \\
    Martensitic interfaces & -24.7 & $5.06 \times 10^{25}$ \\
    \hline
    \end{tabular}
  \label{tab:energy}
\end{table}

\subsection{Numerical model}
\label{Subsec:FE}

Fatigue crack growth rates are estimated by means of a finite element model that combines a gradient-enhanced description of material deformation, hydrogen transport accounting for multiple trap types, and a cohesive zone model sensitive to both cyclic loading and hydrogen degradation (see Section \ref{Sec:Theory}). These three ingredients are implemented in the commercial finite element package \texttt{ABAQUS} by means of user subroutines. Specifically, mass transport is modelled using a \texttt{UMATHT} subroutine that exploits the analogy with heat transfer \cite{Barrera2016,Diaz2016b,EFM2017}, the first-order version of MSG plasticity is implemented by means of a \texttt{UMAT} subroutine \cite{IJSS2015}, and the cohesive zone formulation for hydrogen and fatigue damage is implemented through a \texttt{UEL} subroutine. In addition, a \texttt{DISP} subroutine is used to prescribe the boundary conditions and pre- and post-processing is carried out using the software \texttt{Abaqus2Matlab} \cite{AES2017}.\\ 

\begin{figure}[H]
  \makebox[\textwidth][c]{\includegraphics[width=1.1\textwidth]{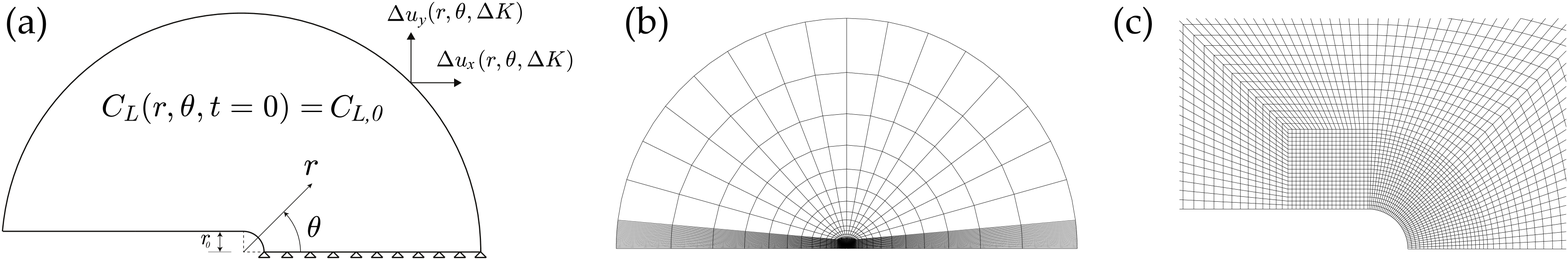}}
  \caption{Boundary value problem: (a) sketch of the boundary layer formulation employed, with mechanical and hydrogen transport boundary conditions, (b) finite element mesh of the entire domain, and (c) finite element mesh of the crack tip region.}
  \label{fig:Boundary}
\end{figure}

Small scale yielding conditions apply, as it is generally the case in fatigue crack growth experiments in the presence of hydrogen, and accordingly predictions are obtained using a boundary layer formulation. As shown in Fig. \ref{fig:Boundary}, a circular region near the crack tip is modelled, where a remote cyclic $K$-field is prescribed by defining the displacement of the nodes located in the outer boundary in agreement with William's \cite{Williams1957} elastic solution. Thus, for a polar coordinate system centered at the crack tip ($r,\theta$), the horizontal and vertical displacements at the outer nodes are given by,
\begin{equation}
   \Delta u_x(r,\theta)=\Delta K\frac{1+\nu}{E}\sqrt{\frac{r}{2\pi}}\cos\left(\frac{\theta}{2}\right)(3-4\nu-\cos\theta) \, ,
    \label{despu}
\end{equation}
\begin{equation}
   \Delta  u_y(r,\theta)= \Delta K\frac{1+\nu}{E}\sqrt{\frac{r}{2\pi}}\sin\left(\frac{\theta}{2}\right)(3-4\nu-\cos\theta) \, .
    \label{despv}
\end{equation}

\noindent Here, $\Delta K$ denotes the mode I stress intensity factor load range: $\Delta K=K_{max}-K_{min}$. Unless otherwise stated, a load range of $\Delta K = 35$ MPa$\sqrt{\text{m}}$ is used, with a load ratio of $R=K_{min}/K_{max} = 0.1$ and a load frequency of $f=1$ Hz. As shown in Fig. \ref{fig:Boundary}, following the works by McMeeking \cite{McMeeking1977a} and Sofronis and McMeeking \cite{Sofronis1989}, an initial blunting radius of $r_0=0.5$ \si{\micro\metre} is defined, while the outer radius is chosen to be sufficiently large to not influence the results ($ > 300,000 r_0$). Taking advantage of symmetry, only one half of the boundary layer circle is simulated, with symmetry boundary conditions being prescribed in the crack ligament.

The crack growth behaviour is reported normalised by a reference stress intensity factor $K_0$, defined as
\begin{equation}
   K_0=\sqrt{\frac{E\phi_n}{1-\nu^2}} \, .
    \label{referenceK}
\end{equation}

\noindent The loading range equals $\Delta K/K_0=0.1$, unless otherwise stated.\\

Regarding hydrogen transport, no hydrogen concentration is prescribed on the boundaries of the domain (Neumann boundary conditions). At time $t=0$, an initial lattice hydrogen concentration $C_{L,0}$ is defined in the entire domain. Plane strain conditions are assumed and the model is discretised with 12,359 eight-node quadrilateral elements with reduced integration. The mesh is refined along the crack propagation region, with the characteristic cohesive length $\delta_n$ being in all cases more than 5 times larger than the characteristic element size, which is sufficient to ensure mesh convergence \cite{EFM2019}. Specifically, we consider $\delta_n=0.075$ mm, using 0.0375 mm in the finite element model due to symmetry, but the results are presented in a normalised fashion and should thus hold for any choice of $\delta_n$. The cohesive elements employed are quadratic with 6 nodes and 12 integration points.

\section{Results}
\label{Sec:Results}

The theoretical and numerical models described in Sections \ref{Sec:Theory} and \ref{Sec:Met}, respectively, are used to study the role of trapping in the fatigue crack growth resistance of metals exposed to an environment containing hydrogen. First, the carbide trap density is varied to investigate the impact of engineering alloys with \emph{beneficial traps} (Section \ref{Sec:TrapDensity}). The sensitivity to the initial hydrogen content is then assessed in Section \ref{sec:InitialH}. Subsequently, in Section \ref{Sec:Frequency}, the role of the loading frequency is quantified. Finally, maps are built to relate the loading frequency and the carbide trap density to the degree of embrittlement, as characterised by an acceleration in fatigue crack growth rates (Section \ref{Sec:Maps}).

\subsection{Influence of carbide trap density}
\label{Sec:TrapDensity}

\begin{figure}[H]
  \makebox[\textwidth][c]{\includegraphics[width=0.8\textwidth]{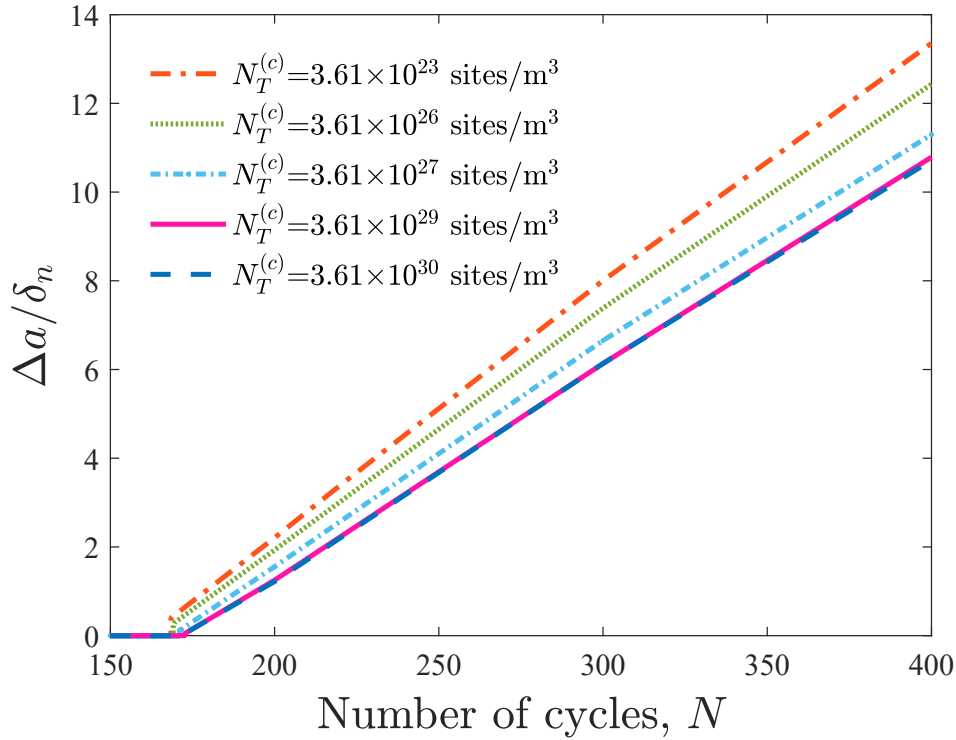}}
  \caption{Influence of the carbide trap density $N_T^{(c)}$ on crack extension versus number of cycles for $\Delta K/K_0$=0.1. Results are obtained under a load ratio of $R=0.1$, a frequency of $f=1$ Hz and an initial lattice hydrogen concentration of $C_{L,0}=1.06$ wt ppm.}
  \label{fig:crecimientof1}
\end{figure}

Fatigue crack growth is simulated for carbide trap densities ranging from $N_T^{(c)}=3.61\times10^{23}$ sites/m$^3$ to $N_T^{(c)}=3.61\times10^{30}$ sites/m$^3$. The results obtained are shown in Fig. \ref{fig:crecimientof1} in terms of the normalised crack extension $\Delta a/ \delta_n$ versus the number of cycles $N$. In all cases, a linear behaviour is observed, with fatigue crack growth rates $da/dN$ decreasing with increasing carbide density $N_T^{(c)}$. This is rationalised as follows. First, note that the lattice hydrogen distribution mimics that of the hydrostatic stress for sufficiently long loading cycles, see Eq. (\ref{Dif}). This leads to an accumulation of hydrogen near cracks and other stress concentrators, with the hydrogen content predicted being 2.5 to 20 times the initial one (depending on whether gradient effects are accounted for \cite{Sofronis1989,CS2020b}). However, if the loading cycle is short (or the material diffusivity is low), then there is less time for hydrogen to diffuse to regions of high $\sigma_H$ and, as a result, lower $C_L$ levels are attained in the fracture region. The limiting cases are the steady state conditions (long loading cycles), where $C_L=C_{L,0}\exp ( \sigma_H \bar{V}_H/ (\mathcal{R}T))$, and very short loading cycles (high $f$) where $C_L=C_{L,0}$. Since, as discussed above, increasing the trap density results in a lower effective diffusivity ($D_e$) and a larger $f/D_e$ ratio, the maximum value of $C_L$ ahead of the crack decreases with increasing $N_T^{(c)}$. Thus, for a given loading frequency, increasing the trap density is a suitable strategy to reduce the impact that hydrostatic stresses have in raising the hydrogen content, resulting in less $C_L$ in the fracture process zone and lower fatigue crack growth rates.

\begin{figure}[H]
  \makebox[\textwidth][c]{\includegraphics[width=0.8\textwidth]{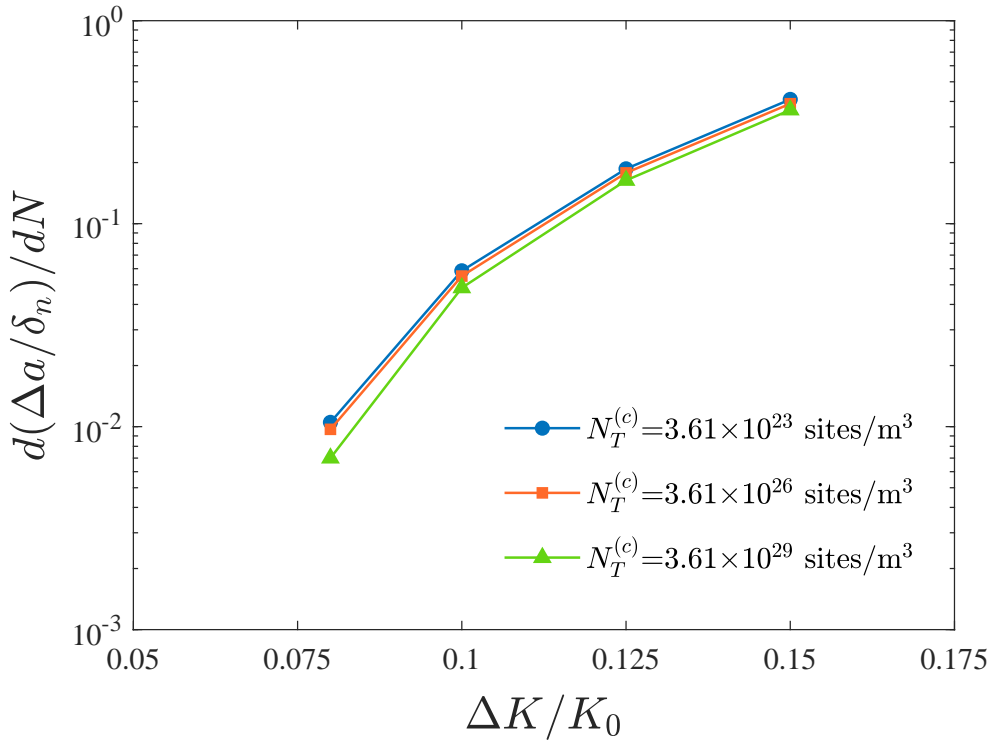}}
  \caption{Influence of the carbide trap density on the fatigue crack growth rates versus load range behaviour. Results are obtained under a load ratio of $R=0.1$, a frequency of $f=1$ Hz and an initial lattice hydrogen concentration of $C_{L,0}=1.06$ wt ppm.}
  \label{fig:dadNf1}
\end{figure}

The results of Fig. \ref{fig:crecimientof1} also show that there is a threshold value of the carbide trap density above which further increases in the magnitude of $N_T^{(c)}$ have no influence in the results. This saturation stage is reached when the ratio $f/D_e$ is sufficiently large and further reducing the diffusion of hydrogen within each cycle has a negligible effect. We observe that, for the conditions analysed here, a magnitude of $N_T^{(c)}=3.61 \times 10^{29}$ sites/m$^3$ is sufficient to reach such a threshold. At this stage, it is important to emphasise that our modelling framework is built upon the assumption of a trap density significantly smaller than the number of lattice sites per unit volume ($N_T << N_L$). Hence, while they are provided here for completeness, results reported for trap densities beyond $1 \times 10^{29}$ are within a regime where accuracy is compromised by modelling assumptions. By running calculations with different load ranges $\Delta K$ and computing the slope of the $\Delta a/ \delta_n$ vs $N$ curve, the influence of the carbide trap density on the Paris law coefficients can be quantified, as shown in Fig. \ref{fig:dadNf1}. 

To quantify the role of the trap density on the Paris law parameters, we consider the data points obtained in the regime $\Delta K/K_0 \geq 0.1$, where the behaviour is linear. Then, the Paris coefficients are fitted by using the following normalised version of the Paris law equation:
\begin{equation}
   \frac{d \left( a/\delta_n\right)}{dN}=C^* \left( \frac{\Delta K}{K_0} \right)^m \, .
    \label{Paris}
\end{equation}

The values obtained for the coefficients $C^*$ and $m$ are provided in Table \ref{tab:Paris} as a function of the trap density of carbides $N_T^{(c)}$. It is shown that $C^*$ increases with carbide density but that the exponent $m$ is fairly insensitive and remains in all cases within the range of experimentally reported values for metals in inert environments ($m \approx 4$). While data on the sensitivity of Paris law coefficients to the trap density has not been reported yet, the results are consistent with experimental and computational data as a function of the hydrogen content \cite{SanMarchi2012,EFM2017}, showing that the influence is significantly more significant in the coefficient $C^*$.

\begin{table}[H]
  \centering
  \caption{Paris law parameters calculated for $da/dN$ [mm/cycle] and $\Delta K$ [MPa$\sqrt m$] depending on the carbide trap density considered.}
    \begin{tabular}{|c|c|c|}
    \hline
    \textbf{$N_T^{(c)}$ [sites/m$^{3}$]} & \textbf{$C^*$} & \multicolumn{1}{c|}{\textbf{$m$}} \\
    \hline
    $3.61 \times 10^{23}$ & 3876.4  & 4.81 \\
    $3.61 \times 10^{26}$ & 3901.2  & 4.84 \\
    $3.61 \times 10^{29}$ & 4917.4  & 4.99 \\
    \hline
    \end{tabular}
  \label{tab:Paris}
\end{table}

\subsection{Influence of the initial lattice hydrogen concentration}
\label{sec:InitialH}

\begin{figure}[H]
  \makebox[\textwidth][c]{\includegraphics[width=0.8\textwidth]{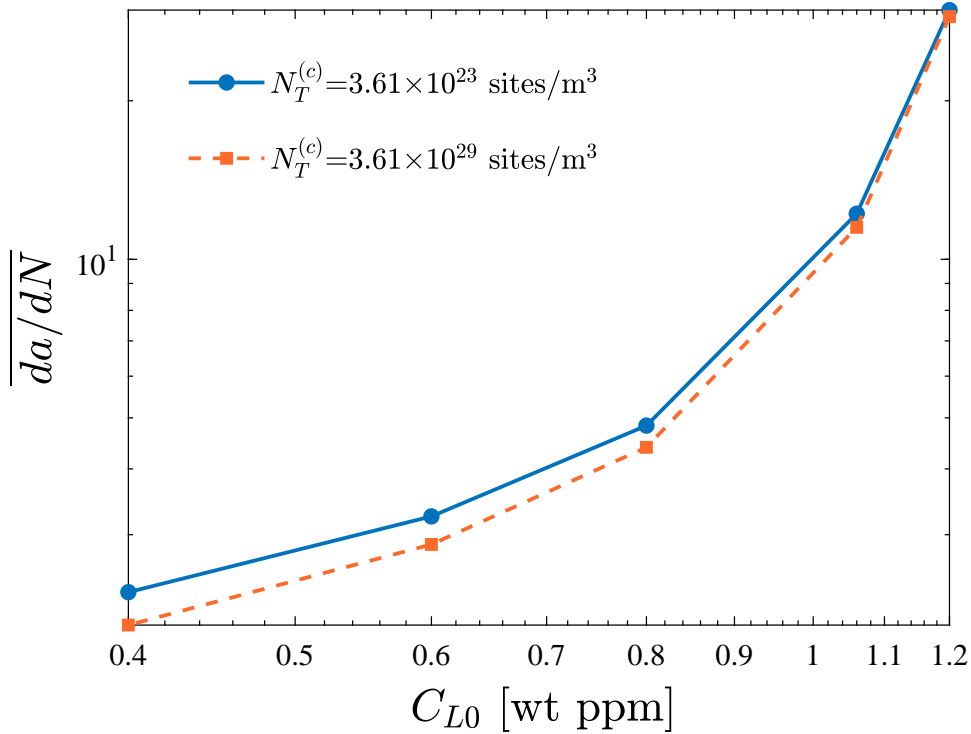}}
  \caption{Influence of the initial lattice hydrogen concentration and the trap density on fatigue crack growth rates. The estimated fatigue crack growth rates have been normalised by the one obtained in the absence of hydrogen; Eq. (\ref{eq:da/dN_normalisation}). Results are obtained under a load ratio of $R=0.1$, a frequency of $f=1$ Hz and $\Delta K/K_0$=0.1.}
  \label{fig:CL0}
\end{figure}

We proceed to evaluate the influence of the initial lattice hydrogen content, $C_{L,0}$. As detailed above, our reference choice ($C_{L0}=1.06$ wt ppm) is based on an estimate of the lattice hydrogen concentration present in the fracture process zone shortly before the onset of crack growth in the experiments \cite{AM2020}. Calculations are conducted within the range $C_{L0}=$0.4-1.2 wt ppm to investigate its impact on hydrogen-assisted fatigue crack growth. The results are reported in Fig. \ref{fig:CL0} in terms of the following normalised fatigue crack growth rate,
\begin{equation}\label{eq:da/dN_normalisation}
    \overline{da/dN} = \frac{\left[ d(\Delta a/\delta_n)/dN \right]_{C_L}}{\left[ d(\Delta a/\delta_n)/dN \right]_{C_L=0}} \, 
\end{equation}

\noindent where the estimated fatigue crack growth rate is divided by the one obtained in the absence of hydrogen ($\left[ d(\Delta a/\delta_n)/dN \right]_{C_L=0}$). Two carbide trap densities are considered, the one measured in the experiments ($N_T^{(c)}=3.61 \times 10^{23}$ sites/m$^3$) and the saturation one, beyond which no influence is observed ($N_T^{(c)}=3.61 \times 10^{29}$ sites/m$^3$).

As expected, and consistent with the phenomenological degradation law adopted (see Fig. \ref{fig:LeyH}), fatigue crack growth rates monotonically increase with the initial hydrogen content. The impact of increasing the trap density is more noticeable for low $C_{L,0}$ values. The trap occupancy $\theta_T$ increases with $C_L$, see Eq. (\ref{eq:Oriani}), and the sensitivity of $D_e$ to an increased trap density is greater when $\theta_T$ is smaller, as per Eq. (\ref{eq:De}). Since the relation between $C_L$ and $\theta_T$ is highly sensitive to the trap binding energy, this also implies that increasing the density of the traps with highest $|W_B|$ would be the most suitable strategy to reduce the maximum levels of $C_L$. Fig. \ref{fig:LeyH} also shows that, for our choice of $C_{L,0}=1.06$ wt ppm, fatigue crack growth rates are roughly 10 times higher than those seen in the absence of hydrogen. However, the experiments show that for a loading frequency of $f=1$ Hz, as considered here, a similar response is predicted for pre-charged and non-charged samples, with embrittlement only observed for smaller frequencies \cite{AM2020}. A frequency of $f=0.1$ Hz is needed to increase crack growth rates by an order of magnitude \cite{AM2020}. Hence, this suggests that the phenomenological cohesive law for AISI 4135 steel adopted (Fig. \ref{fig:LeyH}) is not the most suitable choice to reproduce the hydrogen-assisted fatigue behaviour of the AISI 4140 steel tested. The results from Fig. \ref{fig:CL0} suggest that a more suitable degradation law for the cohesive strength $\sigma_{max,C}$ would be one where no reduction in $\sigma_{max,C}$ is observed until a threshold hydrogen concentration is reached. Unfortunately, to the best of the authors' knowledge, tensile strength versus hydrogen content data for the AISI 4140 steel under consideration have not been reported yet. 

\subsection{Influence of the frequency}
\label{Sec:Frequency}

\begin{figure}[H]
  \makebox[\textwidth][c]{\includegraphics[width=0.8\textwidth]{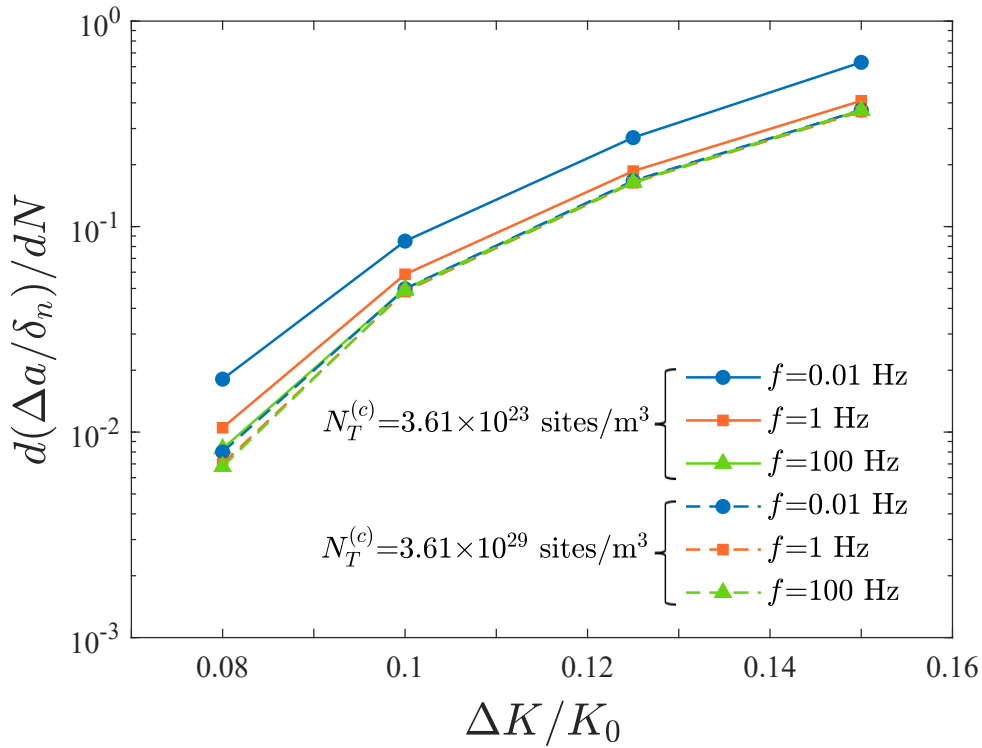}}
  \caption{Influence of the loading frequency and the trap density on the fatigue crack growth rates versus load range behaviour. Results are obtained under a load ratio of $R=0.1$, a frequency of $f=1$ Hz and an initial lattice hydrogen concentration equals to $C_{L0}=1.06$ wt ppm.}
  \label{fig:fPocasMuchas}
\end{figure}

Frequency has an important effect on hydrogen assisted fatigue. Higher fatigue crack growth rates are observed when the loading frequency $f$ is reduced, as the duration of each load cycle is sufficiently large to allow for hydrogen to diffuse and accumulate in the fracture region. We vary the frequency from 0.01 to 100 Hz and compute fatigue crack growth rates for two carbide trap density $N_T^{(c)}$ scenarios: the reference one ($3.61 \times 10^{23}$ sites/m$^3$) and the saturation one ($3.61 \times 10^{29}$ sites/m$^3$). The results are shown in Fig. \ref{fig:fPocasMuchas}.

\begin{figure}[H]
  \makebox[\textwidth][c]{\includegraphics[width=0.8\textwidth]{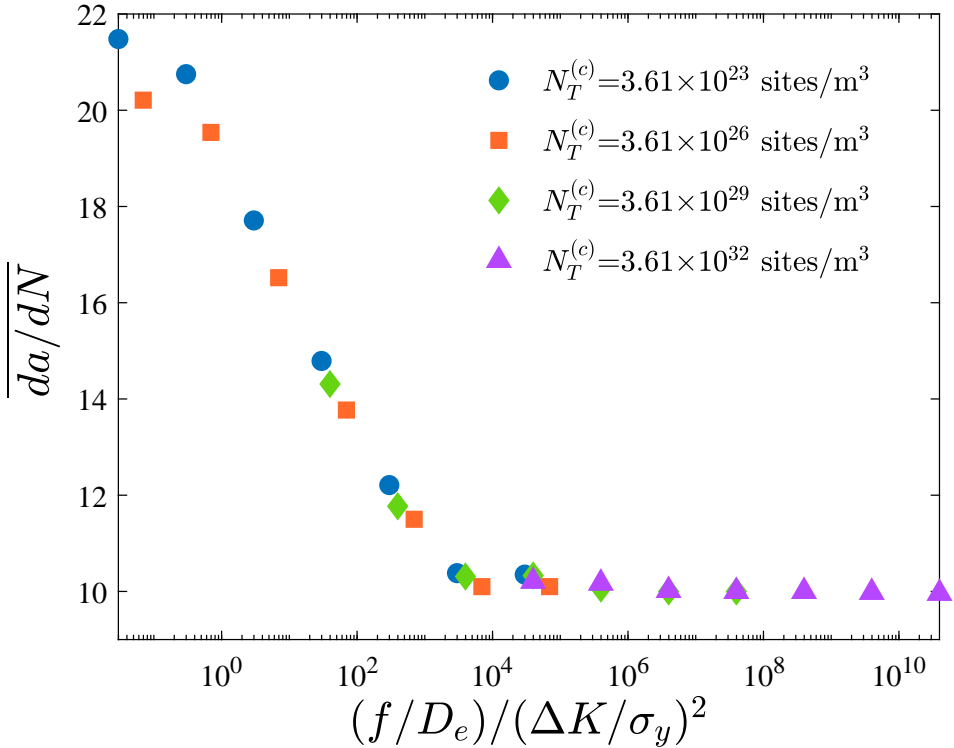}}
  \caption{Influence of the ratio frequency to effective diffusivity ratio and the trap density on fatigue crack growth rates. The estimated fatigue crack growth rates have been normalised by the one obtained in the absence of hydrogen, Eq. (\ref{eq:da/dN_normalisation}), while the ratio $f/D_e$ has been normalised by the stress intensity factor range and the material yield stress. Here $D_e$ corresponds with the initial estimate ($t=0$), without taking into account the evolution of the dislocation density. Results are obtained under a load ratio of $R=0.1$ and $\Delta K/K_0=0.1$.}
  \label{fig:v_f_norm}
\end{figure}

The results shown in Fig. \ref{fig:fPocasMuchas} reveal the expected trend: the smaller the magnitude of $f$, the greater the embrittlement. The comparison between the results obtained with different carbide trap densities shows how the sensitivity to the loading frequency noticeably diminishes for the case of highest $N_T^{(c)}$. While some sensitivity to the frequency is observed for $N_T^{(c)}=3.61 \times 10^{29}$ sites/m$^3$, differences are rather small as the frequency is only decreased by two orders of magnitude and the ratio $f/D_e$ remains comparatively large. Consistently, it can also be observed that the differences between the two $N_T^{(c)}$ scenarios diminish as the loading frequency increases and the ratio $f/D_e$ becomes sufficiently large. The way the degree of susceptibility to hydrogen assisted fatigue is governed by the ratio of loading frequency to effective diffusivity is shown in Fig. \ref{fig:v_f_norm}, where normalised crack growth rates are plotted as a function of a normalised $f/D_e$ and the trap density. 

Fig. \ref{fig:v_f_norm} shows how the results obtained for different carbide trap densities collapse into a single curve when appropriately normalised. Crack growth rates go from 22 times larger than those reported in the absence of hydrogen to 10 times larger in the saturation regime, where $f/D_e$ is sufficiently large. It can observed how the regime of less susceptibility can be reached for all $N_T^{(c)}$ cases considered if the loading frequency is changed accordingly. Or, alternatively, how materials with large densities of beneficial traps remain in the lowest end of fatigue crack growth rates unless exposed to very low loading frequencies. 

\subsection{Mapping the regimes of relevance}
\label{Sec:Maps}

\begin{figure}[H]
  \makebox[\textwidth][c]{\includegraphics[width=0.8\textwidth]{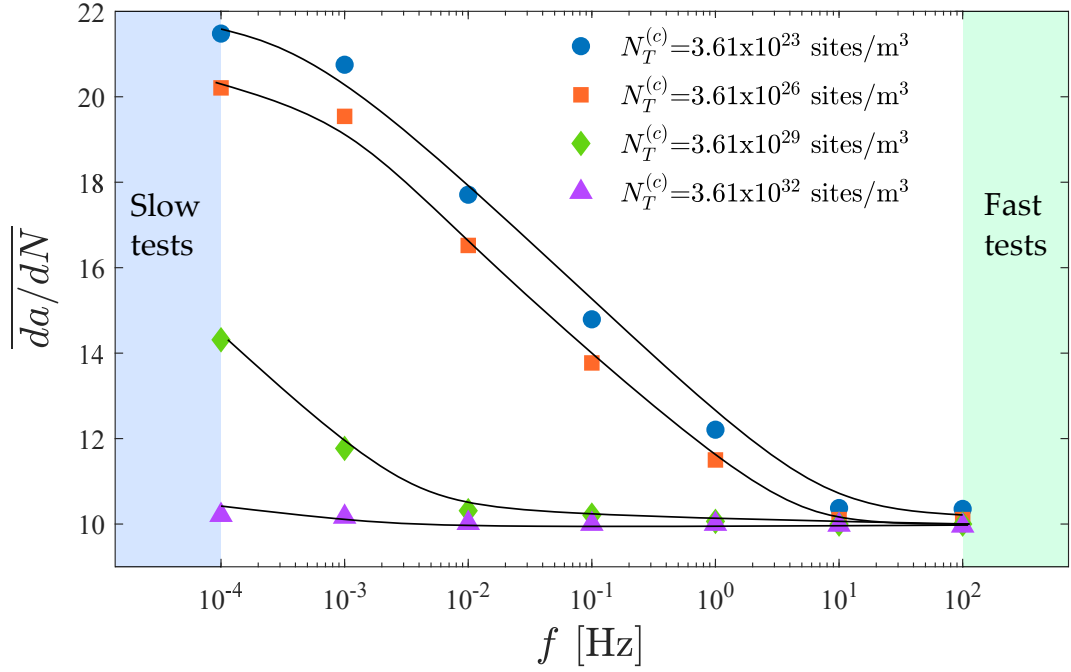}}
  \caption{Mapping the frequency regimes. Fatigue crack growth rate versus frequency for different carbide trap densities. Results have been obtained under a load ratio of $R=0.1$, a load range of $\Delta K/K_0=0.1$ and an initial lattice hydrogen concentration $C_{L0}=1.06$ wt ppm.}
  \label{fig:Vvsf}
\end{figure}

We conclude the Results Section by mapping the regimes of susceptibility as a function of loading frequency $f$ and carbide trap density $N_T^{(c)}$. The goal is to facilitate the design of hydrogen-resistant alloys that exploit the concept of beneficial traps. The two limiting cases in terms of loading frequency regimes can be seen in Fig. \ref{fig:Vvsf}, where normalised fatigue crack growth rates are plotted as a function of $f$ and $N_T^{(c)}$. On the one hand, for sufficiently low frequencies, there is enough time for the hydrogen to diffuse and follow the $\sigma_H$ distribution within each cycle; as a result, hydrogen susceptibility is maximised. For the range of loading frequencies considered in Fig. \ref{fig:Vvsf}, this regime is only achieved for the materials with smallest carbide trap densities ($N_T^{(c)}$ equal to $3.61 \times 10^{23}$ and $3.61 \times 10^{26}$ sites/m$^3$). On the other hand, if the loading frequency is sufficiently large, the diffusion of hydrogen within each loading cycle is negligible and fatigue crack growth rates are at the lower end. It is interesting to note that this requires frequencies of 10 Hz or higher for the original material but that all relevant values of $f$ lie within the regime of lowest susceptibility if the carbide trap density is increased to $3.61 \times 10^{29}$ sites/m$^3$. Another interesting observation is that fatigue crack growth rates duplicate when going from one regime to the other. However, crack growth rates attained at low frequencies are typically reported to be 5 to 10 times larger than those measured at high frequencies \cite{Murakami2010a,EFM2017,AM2020}. These differences with experiments are due to the phenomenological cohesive strength degradation law adopted. As shown in Fig. \ref{fig:LeyH}, the behaviour of the high strength AISI 4135 steel tested by Wang \textit{et al.} \cite{Wang2007} exhibits a particularly significant susceptibility. Together with the piecewise linear fit adopted, this implies that any non-zero hydrogen content will lead to a noticeable drop in the cohesive strength, while no hydrogen susceptibility is typically observed until a threshold content is reached.

\begin{figure}[H]
  \makebox[\textwidth][c]{\includegraphics[width=0.9\textwidth]{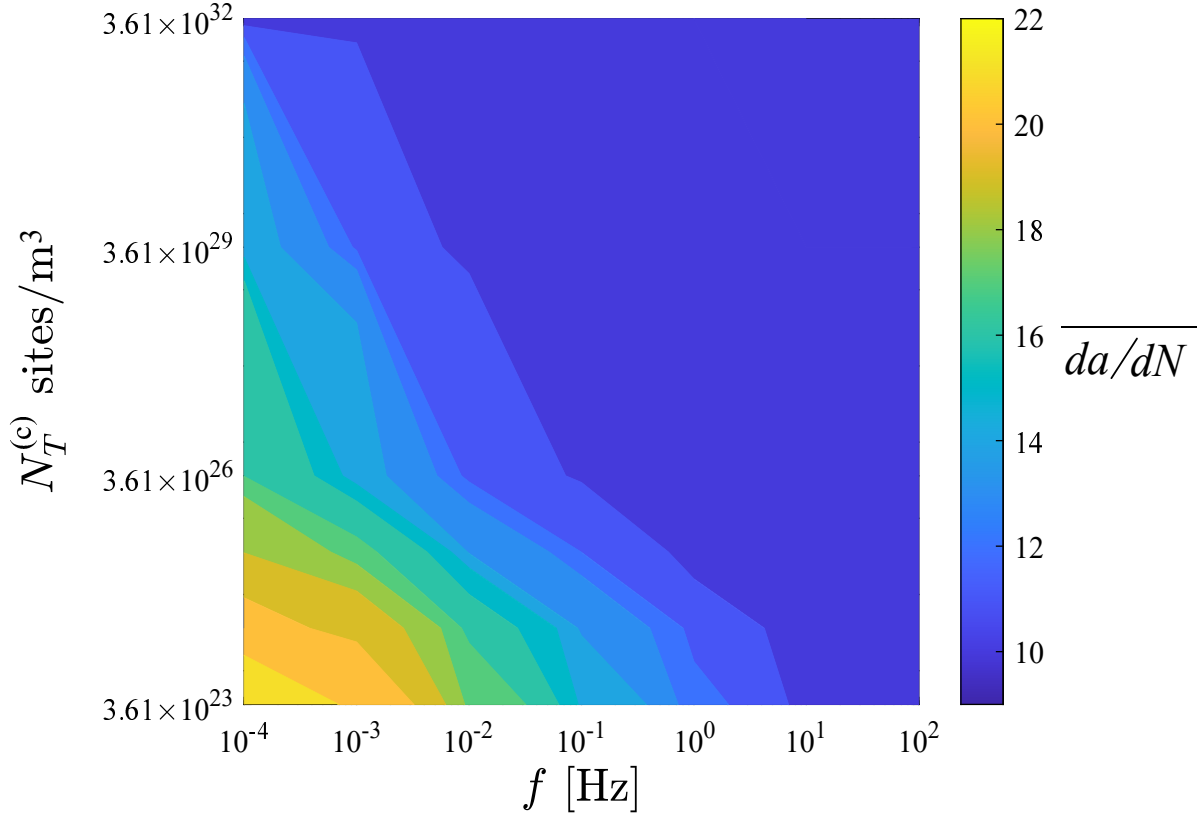}}
  \caption{Mapping the role of trap densities on fatigue crack growth rates as a function of loading frequency $f$. Results have been obtained under a load ratio of $R=0.1$, a load range of $\Delta K/K_0=0.10$ and an initial lattice hydrogen concentration equals to $C_{L0}=1.06$ wt ppm.}
  \label{fig:NT_f_norm}
\end{figure}

Finally, we map the influence of carbide trap densities on hydrogen-assisted fatigue crack growth rates over a wide range of loading frequencies - see Fig. \ref{fig:NT_f_norm}. Let us assume, as seen in the experiments, that the high frequency regime ($f\geq 1$ Hz) leads to hydrogen levels that are insufficient to trigger embrittlement \cite{AM2020}, even for the lowest $N_T^{(c)}$. Then, the blue regions in Fig. \ref{fig:NT_f_norm} denote the regimes where the susceptibility to cracking has been suppressed due to a sufficiently large $f/D_e$ ratio. We can see that increasing the carbide trap density up to $N_T^{(c)}=3.61 \times 10^{32}$ sites/m$^3$ would remove embrittlement effects for the entire range of loading frequencies (up to frequencies as low as $f=0.0001$ Hz). In fact, we only need to increase $N_T^{(c)}$ by three orders of magnitude, something that can be readily achieved, to hinder embrittlement within a technologically-relevant range of loading frequencies (0.1 to 100 Hz). 

The results reported are based upon the assumption that carbides are the trap type whose density is varied to explore the ``beneficial trap'' paradigm, as this has been the strategy followed in the literature so far (see, e.g., \cite{Turk2018}). The impact of trap density on fatigue crack growth rates will be more significant for trap types with larger binding energies. Also, we emphasise that the results presented here employ a theoretical framework that is only valid for $N_T << N_L$ \cite{Krom2000}, and thus results obtained for trap densities above $10^{29}$ sites/m$^3$ should be treated with care.

\section{Conclusions}
\label{Sec:ConcludingRemarks}

We have investigated the role of microstructural traps in hydrogen-assisted fatigue crack growth. To achieve this, a new formulation has been presented, which combines a stress-assisted hydrogen transport model that accounts for multiple traps, an enriched, gradient-enhanced description of crack tip stresses, and a cohesive zone model with cyclic loading and hydrogen degradation effects. The model is particularised to the analysis of a well-characterised AISI 4140 steel with dislocations, carbides and martensitic interfaces as main trap types. Fatigue crack growth rates and Paris law parameters are computed to investigate the influence of critical parameters such as the loading frequency $f$, the initial hydrogen content, and the carbide trap density $N_T^{(c)}$. In the literature, materials with an increased carbide density have been engineered to develop an in-built resilience to hydrogen embrittlement, so varying $N_T^{(c)}$ enables assessing the feasibility of this approach. Our main findings are:

\begin{itemize}
    \item The fatigue crack growth behaviour of metals exposed to hydrogen is governed by the ratio between loading frequency and effective diffusivity ($f/D_e$). The limiting cases are given by sufficiently low $f/D_e$ values, where hydrogen has time to accumulate in areas of high hydrostatic stress, and sufficiently high $f/D_e$ values, where the diffusion of hydrogen within each load cycle is neglible.
    
    \item Increasing the density of carbide traps ($N_T^{(c)}$) diminishes fatigue crack growth rates, due to their influence on the effective diffusivity of the material. A threshold value of $N_T^{(c)}$ exists above which the impact on diffusion is negligible (the ratio $f/D_e$ is sufficiently high). 
    
    \item In terms of Paris law parameters, increasing the trap density has a noticeable effect on the pre-factor but only changes minimally the exponent.
    
    \item The role of traps in reducing fatigue crack growth rates is more significant for lower levels of the initial lattice hydrogen concentration, as the trap occupancy is smaller.
    
    \item For the material and conditions considered, increasing the carbide trap density by three orders of magnitude is sufficient to extend the regime of low susceptibility from frequencies larger than 10 Hz to $f \leq $0.1 Hz. 
    
\end{itemize}

Maps have been provided that enable identifying the combinations of loading frequencies and carbide trap densities that result in a change from high to low hydrogen embrittlement susceptibility. These transitional parameters set the basis for the rational design of alloys with in-built resistance to hydrogen assisted fatigue. 

\section{Acknowledgements}
\label{Sec:Acknowledgeoffunding}

The authors acknowledge funding from the Regional Government of Asturias (grant FC-GRUPIN-IDI/2018/000134) and the IUTA (grant SV-19-GIJON-1-19). E. Mart\'{\i}nez-Pa\~neda was supported by an UKRI Future Leaders Fellowship (grant MR/V024124/1).

%\appendix

%\section{Additional details of numerical implementation}
%\label{App:FEM}

% Appendix A



\bibliographystyle{elsarticle-num}
%{\footnotesize
%\bibliography{library}} % To make the references smaller
\bibliography{library}

%% Text of bib
\end{document}